\documentstyle[12pt]{article}

\begin{document}

\def\NPB#1#2#3{{Nucl.~Phys.} {\bf{B#1}} (19#2) #3}
\def\PLB#1#2#3{{Phys.~Lett.} {\bf{B#1}} (19#2) #3}
\def\PRD#1#2#3{{Phys.~Rev.} {\bf{D#1}} (19#2) #3}
\def\PRL#1#2#3{{Phys.~Rev.~Lett.} {\bf#1} (19#2) #3}

\newcommand{\siml}{\raise -2.truept\hbox{\rlap{\hbox{$\sim$}}\raise5.truept
\hbox{$<$}\ }}
\newcommand{\simg}{\raise -2.truept\hbox{\rlap{\hbox{$\sim$}}\raise5.truept
\hbox{$>$}\ }}

\begin{titlepage}
\vspace{3cm}
\begin{flushright}
UT-809 \\
TU-538, RCNS-98-03 \\
March 1998
\end{flushright}
\vskip 0.8cm
\begin{center}
{\Large \bf Axion Cosmology with its Scalar Superpartner}
\end{center}
\vskip 1.2cm
\begin{center}
M. Hashimoto\renewcommand{\thefootnote}{\fnsymbol{footnote}}%
\footnote{Research Fellow of the
Japan Society for the Promotion of Science.}%
$^1$,  Izawa K.-I.$^1$,
  M. Yamaguchi$^2$ and T. Yanagida$^1$
\\
\vskip 1.5cm
{\em $^1$Department of Physics, University of Tokyo, \\
Tokyo 113-0033, Japan}\\ 
\vskip 1cm
{\em $^2$Department of Physics, Tohoku University, \\
Sendai 980-8578, Japan}
\end{center}
\vskip 2cm
\thispagestyle{empty}
\begin{abstract}
  In supersymmetric theories, the Peccei-Quinn symmetry has a complex
  extension as a symmetry of the superpotential, so that the scalar
  potential always has an almost flat direction, the dilaton. We
  discuss how coherent oscillation of the dilaton affects axion
  cosmology. We stress that the dilaton decay, if its dominant mode
  is not into axions, releases large entropy at a late
  epoch of the Universe's evolution to dilute axion energy density
  and the upperbound of the decay constant is raised up to about
  $10^{15}$ GeV. The case of the M-theory axion is also discussed.
\end{abstract}
\end{titlepage}

Although the standard model describes interactions of elementary
particles remarkably well, its extension is demanded by some fine
tuning problems in it. One of such fine tuning problems is the strong
CP problem \cite{CP} and another is the Higgs mass hierarchy problem.
So far, the most attractive solution to the latter problem is
supersymmetry \cite{SUSY}, whereas the Peccei-Quinn (PQ) mechanism solves
the
former one in a beautiful way \cite{PQ}.  The standard arguments based on
astrophysics and cosmology constrain the PQ symmetry breaking scale to
lie  between $10^9$ GeV and $10^{13}$ GeV \cite{bound1,bound2}.

In this paper, we shall consider the PQ mechanism in the framework of
supersymmetry. Then the axion forms a supermultiplet. We shall discuss
cosmological effects of the scalar superpartner of the axion and show
that it can drastically change the standard axion cosmology.

In supersymmetry, the PQ $U(1)_{PQ}$ symmetry is extended to its 
complex form  $U(1)_{PQ}^{c}$ as a symmetry of the superpotential
\cite{KOY}.
To illustrate this point,  let us consider a linear realization of the $
U(1)_{PQ}$  symmetry, in which a field transforms as
\begin{equation}
   \phi^i \rightarrow e^{i Q_i \alpha} \phi^i,
\end{equation}
where $\alpha$ is a {\em real} parameter of the $U(1)_{PQ}$
transformation and $Q_i$ the $U(1)_{PQ}$ charge of the field.  The
point is that the superpotential $W(\phi)$, a holomorphic function of
the complex fields $\phi^i$, is invariant under the $U(1)_{PQ}$
transformation with $\alpha$ being now an arbitrary {\em complex}
number.  When $\alpha$ is pure imaginary, the transformation
corresponds to a dilatational one. Under the assumption of unbroken
supersymmetry, this $U(1)_{PQ}^{c}$ symmetry leads to the existence of 
a non-compact flat direction in the scalar potential,
associated with this dilatational transformation.
 We shall call, in this paper, the
field along this flat direction a ``dilaton'' field, which will be 
denoted by $\phi$.  Supersymmetry breaking generates the potential for the
dilaton field and thus the dilaton  mass.  In the
gravity-mediated supersymmetry breaking scenario, the dilaton mass,
$m_{\phi}$, is of the order of the gravitino mass, $m_{3/2}$, which
should be in the TeV range.

Existence of the weakly-interacting scalar field, the dilaton, can
drastically change axion cosmology.  In the early Universe the
coherent mode of the dilaton field is in general displaced from its
true minimum.  As the expansion rate of the Universe becomes
comparable to the dilaton mass $m_{\phi}$, the field starts damped
oscillation.  As we will see, the coherent oscillation of the dilaton and
its
subsequent decay may play important roles in the history of the Universe.

Cosmological implications of the dilaton were discussed in
Ref.~\cite{CCK} for a specific case where the PQ scale is radiatively
generated. In the present paper, we stress that the existence of the
flat direction is a general feature and explore its cosmological
consequences in more general situations.

We begin by discussing the decay rate of the dilaton. The dilaton
generally has a coupling to the axion through 
\begin{equation}
{\cal L}_{eff}=\frac{f}{F_a}\phi \partial ^{\mu} a \partial_{\mu} a,
\label{eq:coupling-to-axion}
\end{equation}
where $F_a$ is the $U(1)_{PQ}$ symmetry breaking scale (or the axion
decay constant) and $f$ is a dimensionless parameter which depends on
a model \cite{ChunLukas}. 
Generically it is of order unity but one can also construct a
model with small $f$. From Eq.~(\ref{eq:coupling-to-axion}), 
it follows that the partial decay width is
\begin{equation}
  \Gamma(\phi \rightarrow a a)
 = \frac{f^2}{32 \pi}
   \frac{m_{\phi}^3}{F_a^2}.
\label{eq:decay-width-to-axion}
\end{equation}

Besides the decay into two axions, the dilaton also decays to
standard-model particles.  First we consider a DFSZ axion model
\cite{DFSZ}, in which the axion couples to quarks in the standard
model at the tree level, whereas the coupling to gluons is suppressed
by a one-loop factor. Since the coupling to a quark is proportional to
its mass, the axion couples most strongly to the third-generation quarks. 
 Explicitly it has the following coupling
\begin{equation}
  {\cal L}_{eff} = i\frac{a}{F_a} 
             \left\{ 
      \frac{2x}{x+x^{-1}}  m_t \bar t \gamma_5 t 
     +\frac{2x^{-1}}{x+x^{-1}} m_b \bar b \gamma_5 b +\ldots \right\},
\end{equation}
where $x=1/\tan \beta$ is the ratio of the vacuum expectation values of 
the two
Higgs scalar fields. The coupling of  the dilaton is given by
\begin{equation}
  {\cal L}_{eff} =\frac{\phi}{F_a} 
             \left\{ 
      \frac{2x}{x+x^{-1}}  m_t \bar t  t 
     +\frac{2x^{-1}}{x+x^{-1}} m_b \bar b  b +\ldots \right\}
\label{eq:interaction-DFSZ}
\end{equation}
as  a consequence of supersymmetry.
Assuming that the channel to the $\bar t t$ pair is open, 
one can calculate the decay rate of the dilaton 
\begin{eqnarray}
    \Gamma(\phi \rightarrow t \bar t, b \bar b) =\frac{3}{8 \pi} m_{\phi} 
   & &\left\{ \left( \frac{m_t}{F_a} \right)^2 
         \left( \frac{2x}{x+x^{-1}} \right)^2
         \left(1- \frac{4 m_t^2}{m^2_{\phi}} \right)^{3/2} \right. 
\nonumber \\
   + & & \left. \left( \frac{m_b}{F_a} \right)^2 
      \left( \frac{2 x^{-1}}{x+x^{-1}} \right)^2 
        \left(1- \frac{4 m_b^2}{m^2_{\phi}} \right)^{3/2}\right\}.
\label{eq:decay-width-to-top}
\end{eqnarray}

Secondly in the case of a hadronic axion model \cite{hadronic}, the
coupling to the standard-model quarks vanishes at tree level.  Among
the standard model particles, the axion dominantly couples to the
gluons at the one-loop order. The interaction is given by
\begin{equation}
 {\cal L}_{eff}= \frac{C g_3^2}{32\pi^2} 
     \left( \frac{a}{F_a} F^a _{\mu \nu} \tilde F^{a \mu \nu}
           + \frac{\phi}{F_a} F^a _{\mu \nu} F^{a \mu \nu} \right),
\label{eq:interaction-hadronic}
\end{equation}
where $g_3$ is the $SU(3)_C$ gauge coupling constant and $C$ is a
model dependent constant of order unity. This yields the partial decay 
width into the two gluons
\begin{equation}
   \Gamma (\phi \rightarrow g g ) =
    \frac{2}{\pi} \left( \frac{C \alpha_3}{8 \pi} \right)^2
      \frac{m_{\phi}^3}{F_a^2}.
\label{eq:decay-width-to-gluon}
\end{equation}

We now consider cosmological effects of the decay of the dilaton.
When the dimensionless parameter $f$ in
Eq.~(\ref{eq:coupling-to-axion}) is of order unity, the decay mode
into the two axions tends to dominate over the other modes
\cite{ChunLukas}. The standard big-bang nucleosynthesis (BBN) gives
the upperbound of the energy density of the axions and the dilatons at
the time of the BBN.  Conservatively it must be less than the energy
density contributed by one neutrino species.

The dilaton field starts damped coherent oscillation when the
expansion rate of the Universe becomes equal to the dilaton mass. The
oscillating dilaton eventually decays into the axion pair. The
produced axions do not get thermalized, but just red-shifted. When $F_a
\siml 10^{15}$ GeV, the life-time of the dilaton is shorter than 1
sec, and thus it decays before the BBN starts. If the energy density of 
the dilaton oscillation dominates the energy density of the Universe at 
its decay, the produced axions then dominate the energy density. This
obviously upsets the standard BBN scenario and thus the
dilaton-oscillation energy must not dominate the energy of the
Universe before its decay. Therefore we will consider the case where the
Universe is radiation dominated during the coherent oscillation. Then
the ratio of the dilaton to the radiation energy density at the start
of the oscillation is related to the same ratio at the time of the
dilaton decay in the following way:
\begin{equation}
   \left( \frac{\rho_{\phi}}{\rho_{r}} \right)_{decay} 
\sim
   \left( \frac{\rho_{\phi}}{\rho_{r}} \right)_{osc}   
   \left( \frac{t_{decay}}{t_{osc}} \right)^{1/2},
\end{equation}
where $t_{osc}$ is the time when the oscillation starts and
$t_{decay}$ is the dilaton life-time. If we write the initial amplitude
of the dilaton field as $\epsilon F_a$ with $\epsilon$ being a
dimensionless parameter, the above reads
\begin{equation}
    \left( \frac{\rho_{\phi}}{\rho_{r}} \right)_{decay}   
\sim
    \frac{F_a^2 \epsilon^2}{M^2} 
    \left( \frac{m_{\phi}}{\Gamma_{\phi}} \right)^{1/2}
 \sim 
     \sqrt{\frac{32 \pi}{f^2}}\frac{F_a^3 \epsilon^2}{M^2 m_{\phi}}. 
\end{equation}
Here $M$ represents the reduced Planck mass $2.4 \times 10^{18}$
GeV. The parameter $\epsilon$ depends on the form of the scalar
potential during inflationary epoch.  Generically it is of order
unity, but one can consider also the case where $\epsilon \sim M/F_a$, 
{\em i.e.} the initial amplitude is of the order of the Planck scale.  In
the approximation of the instantaneous decay of the dilaton, the
dilaton energy density is just converted to the produced axion energy
density. We can evaluate the energy density of the produced axions
$\rho_a$ at the time of the primordial nucleosynthesis (with the
temperature $\sim 1$ MeV) as follows:
\begin{equation}
   \left( \frac{\rho_a}{\rho_r} \right)_{1 MeV}
  \sim  \left( \frac{\rho_{\phi}}{\rho_r} \right)_{decay}
      \left( \frac{g_*(T_D)}{g_*(1 \mbox{MeV})} \right)^{1/3}
  \sim
       \sqrt{\frac{32 \pi}{f^2}}\frac{F_a^3 \epsilon^2}{M^2 m_{\phi}}
      \left( \frac{g_*(T_D)}{g_*(1 \mbox{MeV})} \right)^{1/3},
\end{equation}
where $T_D$ denotes the temperature at the time of dilaton decay.

The observation of the primordial abundance of $^4$He  constrains the
energy density of the produced axions at the BBN epoch to be less
than that of the one neutrino species, which in turn means
\begin{equation}
    \left( \frac{\rho_a}{\rho_r} \right)_{1 MeV} <\frac{7}{43}.
\end{equation}
The above constraint leads to a bound on the axion decay constant
\begin{equation}
   F_a \siml 0.6 \times 10^{13} \mbox{GeV} 
      \cdot f^{1/3} \epsilon^{-2/3} 
      \left( \frac{m_{\phi}}{1 \mbox{TeV}} \right)^{1/3}
\label{eq:bound-from-decay-to-axion}
\end{equation}
It is interesting to note that, for $f$, $\epsilon$ =${\cal O}(1)$,
the upperbound of the axion decay constant
(\ref{eq:bound-from-decay-to-axion}) coincides with the upperbound
coming from the closure limit of the axion energy density due to
misalignment.  Note that in this region of $F_a$ the decay of 
the dilaton occurs before the nucleosynthesis and our analysis here is
self-consistent. 
On the other hand, if
$\epsilon$ is much larger than unity, the above constraint is much
severer than the conventional cosmological upperbound.

When the axion decay constant exceeds $10^{15}$ GeV, the dilaton still
oscillates at 1 sec. At this moment, the ratio of the dilaton energy
density to the radiation energy density is evaluated as
\begin{eqnarray}
   \left( \frac{\rho_{\phi}}{\rho_{r}} \right)_{1 sec} 
& \sim &
   \left( \frac{\rho_{\phi}}{\rho_{r}} \right)_{osc}   
   \left( \frac{1 \mbox{sec}}{t_{osc}} \right)^{1/2} \nonumber  \\
& \sim & 10^{9} \epsilon^2 \
     \left(  \frac{F_a}{10^{16} \mbox{GeV}} \right)^2 
     \left( \frac{m_{\phi}}{1 \mbox{TeV}} \right)^{1/2} 
\end{eqnarray}
The $^4$ He abundance constrains the above to be much smaller than
unity. This in turn requires very small $\epsilon$, which is very
unlikely.

If the coupling of the dilaton to the axions, $f$, is small and the
 decay mode into the axion pair is not dominant, the upperbound
$10^{13}$ GeV for the PQ scale can be relaxed in an interesting
way. Namely if the coherent oscillation of the dilaton field dominates
the energy density of the Universe, then its decay followed by 
thermalization produces entropy, which may dilute the energy
density of the axion oscillation. We shall look into this mechanism
in detail. 

{}From the requirement that the decay of the dilaton does not spoil the BBN
scenario, the reheat temperature after the entropy production, $T_R$,
must be higher than 1 MeV. On the other hand, since the axion coherent
oscillation starts around the QCD phase transition, the reheat
temperature should be less than about 100 MeV in order that the
entropy release can dilute the energy density of the axion
oscillation. Because  $T_R$ is related to the decay width
$\Gamma_{\phi}$ as
\begin{equation}
    T_R \sim 0.5 \sqrt{\Gamma_{\phi} M},
\label{eq:TR-general}
\end{equation}
the requirements on the reheat temperature constrain the axion decay
constant, 
in which the dilution mechanism may work.

We shall first consider the DFSZ axion. In this case, the dominant
decay mode is into the top pair. Note that from
Eqs.~(\ref{eq:decay-width-to-axion}) and
(\ref{eq:decay-width-to-top}), we can evaluate  the branching ratio as
\begin{equation}
\mbox{Br}(\phi \rightarrow a a) \sim \frac{f^2}{12}
 \left( \frac{m_{\phi}}{m_t} \right)^2.
\end{equation}
  The BBN
constraint requires that this should be smaller than 7/43. Thus $f$
should be restricted to
\begin{equation}
  f\siml 0.24 \left( \frac{1 \mbox{TeV}}{m_{\phi}} \right).
\end{equation}

{}From Eqs.~(\ref{eq:decay-width-to-top}) and (\ref{eq:TR-general}) it
follows that 
\begin{equation}
    T_R \sim 1.6 \mbox{MeV} \frac{2 x}{x+x^{-1}}
    \left( \frac{10^{15} \mbox{GeV}}{F_a} \right)  
    \left( \frac{m_{\phi}}{1 \mbox{TeV}} \right)^{1/2},
\label{eq:TR-DFSZ}
\end{equation}
where we have neglected the decay mode to the bottom pair.  From this, the
requirement $1$ MeV $ \siml T_R \siml 100$ MeV leads to a constraint on the

axion decay constant:
\begin{equation}
    10^{13} \mbox{GeV} \frac{2 x}{x+x^{-1}}
    \left( \frac{m_{\phi}}{1 \mbox{TeV}} \right)^{1/2} 
    \siml F_a  \siml
    10^{15}\mbox{GeV} \frac{2 x}{x+x^{-1}}
    \left( \frac{m_{\phi}}{1 \mbox{TeV}} \right)^{1/2}. 
   \label{eq:bound-Fa-from-TR}
\end{equation}

In order for the dilution mechanism to work efficiently, 
the energy density of the dilaton oscillation should dominate over
that of the radiation already when the axion field starts oscillation.
The ratio of the dilaton to radiation energy density at a temperature
$T$ is easily calculated to be
\begin{equation}
  \frac{\rho_{\phi}}{\rho_{r}} \sim 
  \left(\frac{\rho_{\phi}}{\rho_{r}}\right)_{osc} 
  \frac{T_{osc}}{T}
\end{equation}
where $T_{osc}$ is the  temperature when the dilaton
oscillation begins.  From this we find that the temperature at which
the dilaton energy equals the radiation energy is roughly
\begin{equation}
    T \sim  100 \mbox{MeV} 
    \left( \frac{\epsilon F_a}{10^{13} \mbox{GeV}} \right)^2  
      \left( \frac{m_{\phi}}{1 \mbox{TeV}} \right)^{1/2}.
\end{equation}
For the efficient dilution, this  should be of the order of 100 MeV or
more, which leads to
\begin{equation}
      \epsilon F_a \simg 10^{13} \mbox{GeV} 
     \left( \frac{m_{\phi}}{1 \mbox{TeV}} \right)^{-1/4}.
\label{eq:efficient-dilution}
\end{equation}

Keeping the above conditions in mind, let us next consider the  
cosmic abundances of the axion. As we discussed, if $T_R
\siml 100$ MeV, the reheating occurs after the axion field starts coherent
oscillation and thus the entropy production dilutes the axion
abundance. The abundance has been calculated in Ref.~\cite{KMY},
with the result
\begin{equation}
   \Omega_a h^2 \sim 5.3 
       \left( \frac{T_R}{1 \mbox{MeV}} \right)
       \left(\frac{ F_a \theta }{10^{16} \mbox{GeV}}\right)^2,
\label{eq:relic-abundance}
\end{equation}
where $\Omega_a$ is the contribution of the axion energy density to the
density
parameter, $h$ is the Hubble constant in units of $100$ km/s/Mpc, and
$\theta$ 
stands for the misalignment of the axion field ($-\pi<\theta<\pi$). 
Plugging Eq.~(\ref{eq:TR-DFSZ}) into the above equation, we obtain
\begin{equation}
   \Omega_a h^2 \sim 0.05 \theta^2 
    \left( \frac{F_a}{10^{15} \mbox{GeV}} \right)
    \left( \frac{m_{\phi}}{1 \mbox{TeV}} \right)^{1/2}.
\end{equation}
{}From this equation we can conclude that the axion energy density
does not overclose the Universe for $10^{13}$ GeV $ \siml F_a \siml
10^{15}$ GeV, corresponding to the reheat temperature 100 MeV $\simg
T_R \simg1$ MeV (see Eq.~(\ref{eq:bound-Fa-from-TR})), 
and thus the entropy production coming from the
dilaton decay revives the DFSZ axion with the decay constant in the
above range%
\setcounter{footnote}{0}%
\footnote{The condition Eq.~(\ref{eq:efficient-dilution})
is satisfied for $ \epsilon
\simg {\cal O}(1).$}%
. Moreover, with $F_a \sim 10^{15}$ GeV and the maximal
value of $\theta \sim\pi$, the axion will be able to constitute the
dominant component of the dark matter of the Universe.

On the other hand, if $F_a \siml 10^{12}$ GeV and hence $T_R \simg 1$ GeV, 
standard computation for the axion relic abundance should apply, namely
\begin{equation}
    \Omega_a h^2 \sim 0.23 \ \theta^2 
      \left( \frac{F_{a}}{10^{12}\mbox{GeV}} \right)^{1.18}.
\label{eq:relic-abundance-standard}
\end{equation}
This  implies that the axion with $F_a \sim 10^{12} $ GeV also provides a
candidate for the dark matter.
  
Next we turn to the case of the hadronic axion.  Comparing
Eq.~(\ref{eq:decay-width-to-axion}) with
Eq.~(\ref{eq:decay-width-to-gluon}), one finds that the dilaton
dominantly decays into gluons, when
\begin{equation}
    f \siml \frac{C \alpha_3}{3 \pi} = {\cal O}(10^{-2})
\end{equation}
for $C={\cal O}(1)$. When it is achieved, the reheat temperature is
evaluated to be
\begin{equation}
   T_R \sim C \mbox{MeV} 
 \left( \frac{10^{14} \mbox{GeV}}{F_a} \right)
 \left( \frac{m_{\phi}}{1\mbox{TeV}} \right)^{3/2}.
\label{eq:TR-hadronic}                  
\end{equation}
With $C={\cal O}(1)$, the requirement $T_R \simg 1$ MeV reads
\begin{equation}
   F_a \siml  10^{14} \mbox{GeV} 
     \left( \frac{m_{\phi}}{1 \mbox{TeV}} \right)^{3/2}
\end{equation}
Note that when $m_{\phi}$ is around 1 TeV, the bound in the hadronic
axion case is one order of magnitude lower than that in the DFSZ
axion. 

For $F_a \sim 10^{12}$ GeV--$10^{14}$ GeV, the reheat temperature is
about 1--100 MeV and therefore the entropy production by the dilaton
decay dilutes the axion relic abundance\footnote{ For $F_a \sim
  10^{12}$ GeV, Eq.~(\ref{eq:efficient-dilution}) requires a large
  initial amplitude of the dilaton field ($\epsilon \simg {\cal
    O}$(10)) to achieve efficient dilution.}.  Using
Eqs.~(\ref{eq:relic-abundance}) and (\ref{eq:TR-hadronic}), one finds
\begin{equation}
   \Omega_a h^2 \sim 5 \times 10^{-4} \ \theta^2 
   \left( \frac{F_a}{10^{14} \mbox{GeV}} \right) 
   \left( \frac{m_{\phi}}{1 \mbox{TeV}}
\right)^{3/2}.\label{eq:Omega-hadronic}
\end{equation}
This shows that the dilution mechanism makes the hadronic axion
with $F_{a} \sim 10^{12}-10^{14}$ GeV   cosmologically
viable. Note, however, even with the maximal $\theta \sim \pi$,
$\Omega_ah^2$ is  at most $5\times 10^{-3}$, too small to be a
dark matter candidate.

When the $U(1)_{PQ}$ breaking scale is as low as $10^{11}$ GeV, the
reheat temperature after the dilaton decay is 1 GeV.  In this case, as
was discussed for the other model, the dilaton decay does not dilute
the axion energy density and hence
Eq.~(\ref{eq:relic-abundance-standard}) can be used to estimate its
relic abundance. For $F_a \sim 10^{11}$ GeV, the hadronic axion can
marginally be the dark matter of the Universe if the maximum $\theta
\sim \pi$ is taken.

Finally we would like to discuss implications of our analysis to the
M-theory axion. It has been argued \cite{BD1,BD2,Choi} that the bulk
moduli fields, living in the eleven-dimensional bulk, provide axion
candidates in the massless spectrum of the strongly-coupled heterotic
string theory (M-theory) compactified on a Calabi-Yau manifold.  If
the volume of the six-dimensional manifold is sufficiently large, the
contributions of any kind of high energy origins to the axion
potential will be strongly suppressed and the dominant contribution
will come from the QCD anomaly.  If this is the case the axion solves
the strong CP problem.  It can be shown that the coupling of the axion
from the bulk moduli to other fields is suppressed by the
(four-dimensional) Planck mass. This means that the decay constant of
the axion is about $10^{16}$ GeV \cite{ChoiKim}.  This axion is
associated with its scalar superpartner whose mass is of the order of
the gravitino mass. A study of the gaugino condensation scenario in
the strong coupling case shows that the soft masses of gauginos and
scalars are comparable to the gravitino mass \cite{NOY}, and thus it
should be in the weak scale to solve the naturalness problem on the
Higgs mass.
Then  we immediately
encounter a difficulty that the scalar partner of the axion decays
after the BBN commences (see Eqs.~(\ref{eq:decay-width-to-gluon}) and
(\ref{eq:TR-hadronic})), with huge entropy produced, which is
obviously a disaster \cite{Polonyi}. To cure this, we have to invoke
another entropy production mechanism with a higher reheat temperature
($T_R \simg 1$ MeV) to dilute the scalar's energy density.  One might
imagine that faster decays of heavier moduli fields, if exist, could
play this role. However, it is not the case, because these heavier
fields start damped oscillation earlier than the dilaton does and the
energy density is eventually dominated by the oscillation energy of
the problematic scalar field (dilaton). To our knowledge, the only
mechanism which can work is the thermal inflation \cite{LS}.  This
mechanism can efficiently dilute the scalar's energy density to a
harmless level\footnote{Note that the reheat temperature should be
  lower than 100 MeV: otherwise the energy density of the axion
  oscillation would not be diluted.}. Then the only constraint on the
$F_a$ comes from the relic abundance of the axion. Applying
Eq.~(\ref{eq:relic-abundance}) to the present case, we conclude that,
with the lowest allowed reheat temperature $T_R \sim 1$ MeV, the
M-theory axion with $F_a \sim 10^{16}$ GeV will survive the
overclosure constraint if $\theta$ is smaller than 0.3.

In addition to the bulk moduli, there may exist moduli fields which
live on the $E_6$ boundary (boundary moduli).  The boundary moduli has
a coupling suppressed only by the GUT scale, a few times $10^{16}$
GeV. If non-perturbative corrections of the superpotential of the
boundary moduli are suppressed, it will be a dominant component of the
QCD axion.  Its scalar superpartner will get a mass comparable to the
other scalars on the same boundary, {\em e.g.} sleptons and squarks.
Applying again the argument of Ref.~\cite{ChoiKim} to estimate the
axion decay constant, one infers $F_a \sim 10^{13}-10^{14}$ GeV
\cite{BD2}. Repeating our arguments for the hadronic axion case (see
Eq.~(\ref{eq:Omega-hadronic})), we may conclude that the axion
originated from the boundary moduli would be cosmologically viable by
itself. However, the existence of the bulk moduli which seems to be
inevitable again requires the late-time entropy production due to the
thermal inflation. Then the QCD axion energy density will be totally
diluted to a cosmologically negligible level. On the other hand, the
other linear combination of the bulk axion and the boundary axion,
which does not couple to QCD anomaly, will gain a mass from
non-perturbative effects of string theory, such as a world-sheet
instanton effect. It will be possible that it becomes the dark matter
of the Universe after the dilution by the thermal inflation.

To conclude, we have discussed the cosmological implications of the
scalar superpartner of the axion, called the dilaton, in the gravity
mediated supersymmetry breaking scenario.  When the dilaton field
dominantly decays into two axions, we have obtained the constraint on
the axion decay constant from the BBN. On the other hand, when the
dilaton dominantly decays into standard model particles, the decay can
release entropy, which dilutes the energy density of the axion
coherent oscillation.  As a result the cosmological upperbound on the
axion decay constraint put by the closure density of the Universe can
be lifted up to about $10^{15}$ GeV for the DFSZ axion and $10^{14}$
GeV for the hadronic axion, respectively.  In the case of the M-theory
axion, we have argued that the thermal inflation with low reheat
temperature is necessary to dilute the energy density of the axion and
its scalar superpartner.

\section*{Acknowledgments}
The work of M.Y. was  supported in part by 
the Grant--in--Aid for Scientific Research from the Ministry of 
Education, Science and Culture of Japan No.\ 09640333.

\end{document}